\documentclass[final,5p,times,twocolumn]{elsarticle}

\usepackage{algpseudocode}
\usepackage{algorithm}
\usepackage{comment}
\usepackage{amssymb}
\usepackage{amsmath}
\usepackage{color}
\usepackage{graphicx}
\usepackage{listings}

\newcounter{bla}

\usepackage{bm}
\newcommand{\vect}[1]{\boldsymbol{\mathbf{#1}}}
\def\vec#1{\vect{#1}}
\journal{Computer Physics Communications}

\begin{document}

\begin{frontmatter}



\title{DEMocritus: an open source framework for simulating slurry flows using an SDPD-DEM coupled model}


\author[RCAST]{Satori Tsuzuki\corref{CA}}
\cortext[CA] {Corresponding author.\\\textit{E-mail address:} tsuzuki@sim.gsic.titech.ac.jp}
\address[RCAST]{Research Center for Advanced Science and Technology, The University of Tokyo, \\4-6-1, Komaba, Meguro-ku, Tokyo 153-8904, Japan}

\begin{abstract}
This paper provides open-source code that works as a viscometer of particle-based simulations of three-dimensional fluid-particle interaction systems, targetting slurry or suspension flow in chemical engineering. The smoothed dissipative particle dynamics (SDPD), a fluid particle model developed for thermodynamic flow at mesoscale, is combined with the contact model of the discrete element method (DEM). The mechanics of fluid-particle interaction is modeled by a two-way interaction scheme using a drag-force model. We demonstrate our simulation code by several validation tests: simulations of reverse-Poiseuille flow, single-particle sedimentation, and a dam-breaking problem containing rigid particles, in reasonable calculation time. Our new open-source code is beneficial for scientists, researchers, and engineers in computational physics.
\end{abstract}


\end{frontmatter}



{\bf PROGRAM SUMMARY}

\begin{small}
\noindent
{\em Program Title:} DEMocritus~Ver~0.0.1\\
{\em Licensing provisions(please choose one): 
}MIT                                   \\
{\em Programming language:} C++17,~Phython3.0~(for job control scripts) \\
{\em Requirements:} Eigen (C++ Library)\\
{\em Operating system:} Linux                                 \\

\end{small}
\section{Introduction}
Viscosity is a dominant parameter determining the rheological characteristics of fluid mechanics. Needless to say, the accurate measurement of viscosity is important in many fields; e.g., the viscosity of slurry consisting of waters and limestone considerably affects the hydration and hardening process of the slurry becoming cement or concrete~\cite{JIANG2016949}. 
Moreover, the viscosity of the slurry in a drain is a critical factor in the clogging of pipes~\cite{Shkundin1974}.
In any case, measuring viscosity in experiments is difficult and costly to realize, and numerical simulations are therefore promising.

However, it has been a challenging topic to realize accurate slurry simulations because a kind of Lagrangian-based approach is indispensable to reproduce the fluid-particle interaction phenomena. Besides, it is necessary to solve the hydrodynamic Navier-Stokes equations and thermodynamic equations simultaneously, because temperature has critical effects on the viscosity of the fluid.

Establishing the methodology of particle simulations of thermodynamic flow was the first hurdle. Many computational physicists have worked on this problem, and a great breakthrough was achieved by Espanol~\cite{PhysRevE.67.026705}, who succeeded in developing smoothed dissipative particle dynamics (SDPD), which incorporates the dissipative particle dynamics (DPD)~\cite{Espaol1995StatisticalMO} into the framework of the smoothed particle dynamics (SPH)~\cite{gingold1977smoothed, monaghan1992smoothed}.
SDPD made it possible to realize SPH simulation of thermodynamic flow including the stochastic Brownian motion of fluid particles. 
Recently, K. Muller et al. reformulated the original SDPD so that it satisfies the NS equations with angular momentum conservation~\cite{MULLER2015301}~
(they named their version SDPD+a). 
D. Alizadehrad et al. applied SDPD+a to a problem of suspension flow of a red cell membrane~\cite{ALIZADEHRAD2018303}. Because SDPD+a incorporates SPH, it can even be applied to free-surface problems.

On the other hand, the discrete element method (DEM), which was proposed by P.A Cundall~\cite{doi:10.1680/geot.1979.29.1.47}, has been widely accepted as a straightforward method for analyzing contacting rigid particles in broad areas of civil engineering. Studies on combining SPH or MPS~\cite{koshizuka2018moving} with DEM to reproduce fluid-structure interaction (FSI) phenomena~(e.g., sloshing tanks, tsunamis containing rubble), have also been intensively conducted in marine and coastal engineering~\cite{doi:10.1080/21664250.2018.1436243}. 

This paper provides open-source code that works as a viscometer of particle-based simulations of three-dimensional fluid-particle interaction systems, targetting slurry or suspension flow in chemical engineering. We combine SDPD with the contact model of the DEM. The mechanics of fluid-particle interaction is modeled by two-way interaction schemes using a drag-force model. 
Unexpectedly to the author, few related studies have evidently remarked on a way of coupling SDPD with the DEM, despite them both being well-established particle models. As mentioned above, SDPD is the almost same as SPH other than calculating the random forces among fluid particles. 
As the first step of realizing SDPD-DEM simulations, our code introduces a coupling technique of SPH with DEM~\cite{SUN2013147, HE2018548} into that of SDPD with DEM, although further studies on the SDPD-DEM coupling model are still expected. Nevertheless, the current model of SDPD-DEM (an alternative model using SPH-DEM), is experimentally shown to be acceptable at a certain level through benchmark tests (see Section~4).

The remainder of this paper is structured as follows. Section 2 briefly describes the numerical models programmed in our code. Section 3 describes the implementation and design of our framework. In Section 4, we demonstrate our framework in several example cases. Section 5 summarizes our results and concludes the paper.

\section{Numerical models}\label{sec:nmmodels}
Based on the concept of fluid particle modeling~\cite{PhysRevE.57.2930}, the Newtonian dynamics of fluid-particle interaction systems consisting of $N$ particles can be modeled as
\begin{align}
& {\small m_{i}{\frac{d\vec{v}_{i}}{dt}} 
	= \sum_{i \ne j} \vec{F}_{ij}^{\rm SDPD}
	+ \sum_{i \ne j} \vec{F}_{ij}^{\rm DEM} 
	+ \sum_{i \ne j} \vec{F}_{ij}^{\rm LUB} 
	+ \sum_{i \ne j} \vec{F}_{ij}^{\rm FPI}
	+ \vec{F}_{i}^{\rm EXT}}, \label{eq:govern}
\end{align}
where $m_{i}$ and $\vec{v}_{i}$ are the mass and velocity of the $i$th particle, respectively. Subsequently, $\vec{F}_{ij}^{\rm SDPD}$ is a fluid force, $\vec{F}_{ij}^{\rm DEM}$ is a contacting force, and $\vec{F}_{ij}^{\rm LUB}$ is a lubrication force between the $i$th and $j$th particles. Note that the latter two forces are valid only when both particles are rigid. Meanwhile, $\vec{F}_{ij}^{\rm FPI}$ is a fluid-particle interaction force, which is valid only when one is a rigid particle and the other is a fluid particle. 
$\vec{F}_{i}^{\rm EXT}$ is an external force given to the $i$th particle.
In our program, we apply well-known numerical models to compute each force in the right-hand side of Eq.~(\ref{eq:govern}).
The details of the respective forces are explained in the following sections.

\subsection{Fluid forces} \label{seq:fluidforce}
An SPH discretization of the Navier-Stokes equations with spin angular momentum conservations~\cite{doi:10.1063/1.1711295} yields three forces: conservation force, dissipative force, and rotational force~\cite{MULLER2015301, ALIZADEHRAD2018303}. We denote the conservation force as $\vec{F}_{ij}^{C}$, dissipative force as $\vec{F}_{ij}^{D}$, and rotational force as $\vec{F}_{ij}^{R}$. The SDPD+a by K. Muller~\cite{MULLER2015301} reformulates them and combines them with the random force $\vec{F}_{ij}^{T}$ of DPD. Each component of total fluid force $\vec{F}_{ij}^{\rm SDPD}$ is described as
\begin{eqnarray}
\vec{F}_{ij}^{C} &=& \Bigr(\frac{p_{i}}{\rho^{2}_{i}} + \frac{p_{j}}{\rho^{2}_{j}} \Bigl)F_{ij}\vec{r}_{ij}, \label{eq:FijC} \\
\vec{F}_{ij}^{D} &=& -\gamma^{a}_{ij}\Bigr( \vec{v}_{ij} 
		             + \frac{\vec{e}_{ij}(\vec{e}_{ij}\cdot\vec{v}_{ij})}{3} \Bigl)
                     - \frac{2\gamma^{b}_{ij}}{3}\vec{e}_{ij}(\vec{e}_{ij}\cdot\vec{v}_{ij}), \label{eq:FijD} \\
\vec{F}_{ij}^{R} &=& -\gamma^{a}_{ij} \frac{\vec{r}_{ij}}{2}\times(\vec{\omega}_{i}+\vec{\omega}_{j}), \label{eq:FijR} \\
\vec{\tilde{F}}_{ij} &=& \Bigr( \sigma^{a}_{ij} d \vec{\overline{W}}^{S}_{ij} 
		                + \sigma^{b}_{ij}\frac{1}{3}{\rm tr}[d \vec{{W}}_{ij}]\mathbb{1}\Bigl)\cdot\frac{\vec{e}_{ij}}{dt} \label{eq:FijT}.
\end{eqnarray}
Here, $p_{i}$ and $p_{j}$ are the pressures of the $i$th and $j$th particles, respectively~(we discuss pressure calculations later). 
$\rho_{i}$ and $\rho_{j}$ are the densities of the $i$th and $j$th particles.  
Similarly, $\omega_{i}$ and $\omega_{j}$ are the respective spin angular velocities.
The relative vector $\vec{r}_{ij}$, relative velocity $\vec{v}_{ij}$, and unit vector $\vec{e}_{ij}$ are given as
\begin{eqnarray}
\vec{r}_{ij} &:=& \vec{r}_{i}-\vec{r}_{j},  \\
\vec{v}_{ij} &:=& \vec{v}_{i}-\vec{v}_{j},  \label{eq:vijbreak}\\
\vec{e}_{ij} &:=& \frac{ \vec{r}_{ij}}{|\vec{r}_{ij}|}. \label{eq:eijbreak}
\end{eqnarray}
Besides, $F_{ij}$ in Eq.~(\ref{eq:FijC}) is defined using a gradient of a kernel function $W(r)$ as
\begin{eqnarray}
F(|\vec{r}_{ij}|) &:=& \frac{1}{\vec{r}} \nabla W(|\vec{r}_{ij}|). \label{eq:defF}
\end{eqnarray}
There are several candidates of the kernel function $W(r)$ that satisfy Eq.~(\ref{eq:defF}).
In this paper, we choose to use a classical kernel function, the Lucy kernel~\cite{lucy1977numerical}, which is expressed as  
\begin{eqnarray}
W(r) &=& \frac{105}{16 \pi h^3}\Biggl(1+3\frac{r}{h}\Biggr)\Biggl(1-\frac{r}{h}\Biggr)^3. \label{eq:kernel}
\end{eqnarray}
At this time, the function $F(r)$ is given as 
\begin{eqnarray}
F(r) &=& \frac{315}{4\pi h^5}\Biggl(1-\frac{r}{h}\Biggr)^2.  \label{eq:F}
\end{eqnarray}
Meanwhile, the parameters of $\gamma^{a}_{ij}$, $\gamma^{b}_{ij}$, $\sigma^{a, b}_{ij}$, and $d\vec{\overline{W}}^{S}_{ij}$ are given as
\begin{eqnarray}
\gamma^{a}_{ij} &=& \Bigr( \frac{20\eta}{3}-4\xi\Bigl)\frac{F_{ij}}{\rho_{i}\rho_{j}}, \\
\gamma^{b}_{ij} &=& \Bigr( 17\xi-\frac{40\eta}{3}\Bigl)\frac{F_{ij}}{\rho_{i}\rho_{j}}, \\
\sigma^{a,b}_{ij} &=& 2\sqrt{k_{B}T\gamma^{a,b}_{ij}}, \\
d \vec{\overline{W}}^{S}_{ij} &=& \frac{1}{2}(d\vec{W}_{ij} + d\vec{W}_{ij}) - \frac{1}{3}{\rm tr}[d\vec{W}_{ij}]. \label{eq:reldWij}
\end{eqnarray}
Here, $\eta$, and~$\zeta$ are dynamic shear viscosity and bulk viscosity, respectively. $k_{B} T$ is a parameter of energy level of the whole system. $d\vec{W}_{ij}$ is a tensor matrix of independent Wiener increments, each element of which can be expressed as $\sqrt{dt}\zeta$, where $\zeta_{ij}\sim N(0,1)$ is an identical independent Gaussian random variable~\cite{quarteroni2015modeling}. ${\rm tr[\vec{X}]}$ represents the trace of a matrix $\vec{X}$.

Among the four forces, $\vec{F}_{ij}^{C}$, $\vec{F}_{ij}^{D}$, and $\vec{F}_{ij}^{R}$ are derived from the SPH discretization of the Navier-Stokes equations with angular momentum conservation. On the other hand, $\vec{F}_{ij}^{T}$ is derived from the DPD formula. Additionally, the matrix $\rm{d}\vec{W}$ of $\vec{F}_{ij}^{T}$ gives symmetric randomness to each pair of neighboring particles. 

Input parameters needed for computing the fluid force are the shear viscosity $\eta$, bulk viscosity~$\zeta$, a parameter of energy level $k_{B} T$, reference density $\rho_{0}$, and reference pressure $p_{0}$. The density of each particle is computed similarly to SPH at every time step by using 
\begin{eqnarray}
\rho_i &=& \sum_{j} m_{j} W(\vec{r}_{ij}), \label{eq:compdens}
\end{eqnarray}
where $j$ is an index of one of the neighboring particles of the $i$th particle. 
Also, the pressure of each particle is computed using the equation of state for pressure as
\begin{eqnarray}
p_{i} = p_{0}\Biggl(\frac{\rho_{i}}{\rho_{0}}\Biggr)^{\alpha} - \beta, \label{eq:prtcpres}
\end{eqnarray}
where $\rho_0$ and $p_{0}$ are the reference density and reference pressure, respectively, $\alpha$ is a free parameter to control the compressibility of the system, and $\beta$ is a parameter that determines the initial pressure of the system. Furthermore, we introduce the techniques of pressure filtering~\cite{Imoto2019} and density reinitialization~\cite{COLAGROSSI2003448} to maintain the numerical stability.

\subsection{Rigid forces} \label{sec:rigidforce}
The contacting force $\vec{F}_{ij}^{\rm DEM}$ between the $i$th and $j$th particles 
is decomposed into a normal component $\vec{F}_{{n}_{ij}}$ and a tangential component $\vec{F}_{{t}_{ij}}$,
which are represented as follows~\cite{doi:10.1680/geot.1979.29.1.47}:
\begin{eqnarray}
\vec{F}_{{n}_{ij}} &=& -k_{{n}_{ij}}\vec{\delta}_{{n}_{ij}}-c_{{n}_{ij}}\vec{v}_{{n}_{ij}}, \label{eq:dem_fnij} \\
\vec{F}_{{t}_{ij}} &=& -k_{{t}_{ij}}\vec{\delta}_{{t}_{ij}}-c_{{t}_{ij}}\vec{v}_{{t}_{ij}}.
\end{eqnarray}
Here, $k_{{n}_{ij}}$ and $c_{{n}_{ij}}$ represent the spring coefficient and dumping coefficient in the normal direction, respectively.
$k_{{t}_{ij}}$ and $c_{{t}_{ij}}$ represent the spring coefficient and dumping coefficient in the tangential direction, respectively.

The normal velocity $\vec{v}_{{n}_{ij}}$ and normal displacement $\vec{\delta}_{{n}_{ij}}$ are given as 
\begin{eqnarray}
\vec{v}_{{n}_{ij}} &=& [(\vec{v}_{i}-\vec{v}_{j})\cdot \vec{n}_{ij}] \vec{n}_{ij}, \\
\vec{\delta}_{{n}_{ij}} &=& [L_{ij}-(R_{i}+R_{j})] \vec{n}_{ij}.
\end{eqnarray}
Here, $L_{ij}$ is the distance between the $i$th and $j$th particles.
$R_{i}$ and $R_{j}$ are the radii of the $i$th and $j$th particles, respectively.
On the other hand, $\vec{n}_{ij}$ represents a normal vector defined as 
\begin{eqnarray}
 \vec{n}_{ij} &:=& \frac{\vec{x}_{i} - \vec{x}_{j}}{|\vec{x}_{i} - \vec{x}_{j}|}, 
\end{eqnarray}
where $\vec{x}_{i}$ and $\vec{x}_{j}$ are the positions of the $i$th and $j$th particles, respectively.
The tangential velocity $\vec{v}_{{t}_{ij}}$ and tangential displacement $\vec{\delta}_{{t}_{ij}}$ are given as 
\begin{eqnarray}
\vec{v}_{{t}_{ij}} &=& \vec{v}_{ij} - (\vec{v}_{ij}\cdot\vec{n}_{ij})\vec{n}_{ij} + (R_{i}\vec{\omega_{i}} + R_{j}\vec{\omega_{j}})\times\vec{n}_{ij}, \\
\vec{\delta}_{{t}_{ij}} &=& \int_{t_{\alpha}}^{t_{\beta}}\vec{v}_{{t}_{ij}}dt. \label{eq:tandispseq}
\end{eqnarray}
Here, $t_{\alpha}$ and $t_{\beta}$ respectively represent the beginning and end times of the interaction between the $i$th and $j$th particles.
The discretized expression of Eq.~(\ref{eq:tandispseq}) representing the relationship between the displacement of $\vec{\delta}_{{t}_{ij}}^{n}$ at the time step $n$ and that of $\vec{\delta}_{{t}_{ij}}^{n+1}$ at the time step $n+1$ can be described as
\begin{eqnarray}
\vec{\delta}_{{t}_{ij}}^{n+1} &=& |\vec{\delta}_{{t}_{ij}}^{n}|\vec{t}_{ij} + \vec{v}_{{t}_{ij}}^{n} \Delta t, \label{eq:tandispn} 
\end{eqnarray}
where, the normal vector $\vec{t}_{ij}$ in the tangential direction is given as
\begin{eqnarray}
\vec{t}_{ij} &=&
\begin{cases}
\frac{\vec{\delta}_{{t}_{ij}}^{n}}{|\vec{\delta}_{{t}_{ij}}^{n}|} & (\vec{v}_{{t}_{ij}}^{n} = 0), \\
\frac{\vec{v}_{{t}_{ij}}^{n}}{|\vec{v}_{{t}_{ij}}^{n}|} & (\vec{v}_{{t}_{ij}}^{n} \ne 0).
\end{cases}
\end{eqnarray}
The effect of the friction coefficient $\mu$ on the surfaces of rigid particles is modeled as
\begin{eqnarray}
\vec{F}_{{t}_{ij}} &=& -\mu |\vec{F}_{{n}_{ij}}| \vec{t}_{ij}~~~~(|\vec{F}_{{t}_{ij}}| \ge \mu |\vec{F}_{{n}_{ij}}|). \label{eq:slideref}
\end{eqnarray}
Equation~(\ref{eq:slideref}) is called the ``slider-effect''.

The coefficients of the DEM are determined according to the Hertz-Mindlin contacting theory~\cite{fischer2000introduction}. 
Normal spring coefficient $k_{{n}_{ij}}$ is described as 
\begin{eqnarray}
k_{{n}_{ij}} &=& \frac{4}{3\pi} \Biggl(\frac{1}{D_{i}+D_{j}}\Biggr)\sqrt{\frac{R_{i}R_{j}}{R_{i}+R_{j}}}, \label{eq:deminputA} \\
D_{i} &=& \frac{1-{\nu_{i}}^2}{E_{i}\pi}, \label{eq:deminputC} \\
D_{j} &=& \frac{1-{\nu_{j}}^2}{E_{j}\pi}. \label{eq:deminputD}
\end{eqnarray}
Here, $R_x~(x = i,j)$ is the radius, $\nu_x$ is Poisson's ratio, and $E_x$ is Young's modulus of each particle.
The tangential spring coefficient $k_{{t}_{ij}}$ is obtained from the definition of $\rm Lam\acute{e}'$s constants as
\begin{eqnarray}
k_{{t}_{ij}} &=& \frac{k_{{n}_{ij}}}{2(1+\nu_{i})}.\label{eq:deminputB}
\end{eqnarray}
The normal damping coefficient $c_{{n}_{ij}}$ and tangential damping coefficient $c_{{t}_{ij}}$ are obtained 
by solving the equations of the harmonic oscillator of the Kelvin-Voigt model as
\begin{eqnarray}
c_{{n}_{ij}} &=& 2\sqrt{\frac{2 m_{i} m_{j}}{m_{i}+m_{j}}k_{{n}_{ij}}},\label{eq:deminputE} \\
c_{{t}_{ij}} &=& 2\sqrt{\frac{2 m_{i} m_{j}}{m_{i}+m_{j}}k_{{t}_{ij}}},\label{eq:deminputF}
\end{eqnarray}
where $m_x~(x=i,j)$ is the mass of each particle.
To summarize, input parameters of the contacting forces are $R_x$, $m_x$, $\nu_x$, and $E_x$; the spring coefficients of $k_{{n}_{ij}}$ and $k_{{t}_{ij}}$ and damping coefficients of $c_{{n}_{ij}}$  $c_{{t}_{ij}}$ are computed at every pair of contacting particles using Eq.~(\ref{eq:deminputA}) to Eq.~(\ref{eq:deminputF}) in our code.

\subsection{Lubrication forces}
The lubrication force is a long-range interaction force that reproduces the exclusion effect of fluid volume existing between two rigid particles.
According to~~\cite{schwarzkopf2011multiphase}, the lubrication force $F_{ij}^{\rm LUB}$ between the $i$th and $j$th rigid particles can be described as
\begin{eqnarray}
F_{ij}^{\rm LUB} &=& \frac{3\pi \mu_{f} {d^{2}}(\vec{v}_{j} - \vec{v}_{i}) }{8(|\vec{r}_{j} - \vec{r}_{i}| - d)}. \label{eq:lubri} 
\end{eqnarray}
Here, $\mu_{f}$ is the fluid viscosity. 
$\vec{v}_{i}$ and $\vec{v}_{j}$ are the velocities and $\vec{r}_{i}$ and $\vec{r}_{j}$ are the positions of the $i$th and $j$th particles, respectively. 
$d$ is the diameter of a rigid particle. 
In our program, we set a cut-off radius of $F_{ij}^{\rm LUB}$ to be the kernel radius $h$ of SDPD; we only compute $F_{ij}^{\rm LUB}$ among the particles existing inside the kernel radius $h$.

\subsection{Fluid-particle interaction forces}\label{sec:fpiforce}
The fluid-particle interaction force $\vec{F}_{ij}^{\rm FPI}$ that acts from the $j$th fluid particle to the $i$th rigid particle is decomposed into a pressure gradient force and drag force. 
We introduce the ``single pressure hypothesis''~\cite{keyfitz2001mathematical}, which assumes rigid particles have the same pressure as fluid particles. Under this assumption, we compute the particle pressure of rigid particles using Eq.~(\ref{eq:prtcpres}) in the same way as fluid particles. Namely, rigid particles are regarded as fluid particles in the calculation of density and pressure of fluid particles. We compute the pressure gradient force from the fluid to a rigid particle by the weighted mean of the fluid forces of the neighboring particles. To be safe, we ignore the rotational force and random force; the pressure gradient force~($\Delta P$ force) $\vec{F}_{ij}^{\rm PG}$ that acts on the $i$th rigid particle can be described as
\begin{eqnarray}
\vec{F}_{ij}^{\rm PG} &=& \frac{\sum_{j} d \rho_0 (\vec{F}_{ij}^{C} + \vec{F}_{ij}^{D}) W_{ij}}{\sum_{j} W_{ij}}, \label{eq:F_fpi_PG} 
\end{eqnarray}
where $\rho_0$ is the reference density of fluid, and the parameter $d$ is the diameter of a rigid particle.

Meanwhile, drag force $\vec{F}_{ij}^{\rm DG}$ is given by the velocity difference between a rigid particle and the fluid around the rigid particle, the fluid volume fraction $\epsilon$, and the momentum exchange coefficient $\beta$ as follows:
\begin{eqnarray}
\vec{F}_{ij}^{\rm DG} &=& \frac{\beta V_{s}}{1-\epsilon}(\vec{v}_i - \vec{v}_f), \label{eq:F_fpi_DG}
\end{eqnarray}
where $V_{s}$ is the volume of a rigid particle, $\vec{v}_i$ is the velocity of $i$th rigid particle, and $\vec{v}_{f}$ is the velocity of the fluid averaged by the neighboring fluid particles of the rigid particle. Here, the moment exchange coefficient $\beta$ is a semi-empirical parameter. Ergun and Wen-Yu~\cite{doi:10.1021/ie50474a011, 10003815634} experimentally measured the value of $\beta$ to obtain the following relationship: 
\begin{eqnarray}
\beta &=&
\begin{cases}
 150\frac{\mu_{f} (1-\epsilon)^2}{\epsilon d^2} + 1.75\frac{(1-\epsilon)\rho}{d}|\vec{v}_f - \vec{v}_i|& (\epsilon \le 0.8), \\
\frac{3}{4}C_{D} \frac{\epsilon(1-\epsilon)}{d}\rho|\vec{v}_f - \vec{v}_i|\epsilon^{-2.65} & (\epsilon \ge 0.8).
\end{cases}
\label{eq:beta}
\end{eqnarray}
The parameters $\rho$ and $\mu_{f}$ are the density and viscosity of the fluid, respectively. The parameter $d$ represents the diameter of a rigid particle. $C_{D}$ is the drag coefficient, which is given by
\begin{eqnarray}
C_{D} &=&
\begin{cases}
\frac{24(1.0+0.15 Re_{p}^{0.687})}{Re_{p}}& (Re_{p} \le 1000), \\
0.44 & (Re_{p} \ge 1000).
\end{cases}
\end{eqnarray}
where the parameter $Re_{p}$ is defined as
\begin{eqnarray}
Re_{p} := \frac{\rho d \epsilon}{\mu_{f}}|\vec{v}_f - \vec{v}_i|.
\end{eqnarray}
$Re_{p}$ is called the particle Reynolds number.

The fluid-particle interaction force that acts from rigid particles to the $i$th fluid particle is decomposed into a pressure gradient force and a reaction force of the drag force.
Because of the single-pressure hypothesis, the pressure gradient force is included in the calculation of $\vec{F}_{ij}^{\rm SDPD}$. The other component force of the fluid-particle interaction is given by the resultant force of the reactions of $\vec{F}_{ij}^{\rm DG}$ among the neighboring particles.
Because of the conservation of momentum exchange between fluid and rigid particles, the reaction force of Eq.~(\ref{eq:F_fpi_DG}) to the $i$th fluid particle $\vec{F}_{i}^{\rm RXN}$ is obtained by a weighted calculation among the neighboring particles~\cite{SUN2013147, HE2018548} as 
\begin{eqnarray}
\vec{F}_{i}^{\rm RXN} &=& -\frac{\sum_{j} \vec{F}_{ij}^{\rm DG} W(\vec{r}_{ij})}{\sum W(\vec{r}_{ij})}. \label{eq:F_fpi_DG_reac}
\end{eqnarray}
Recall that $\vec{F}_{i}^{\rm EXT}$ includes $\vec{F}_{i}^{\rm RXN}$ in the expression of Eq.~(\ref{eq:govern}).

\subsection{Boundary treatments}\label{seq:BCs}
\begin{figure}[h]
\centerline{\includegraphics[scale=0.25]{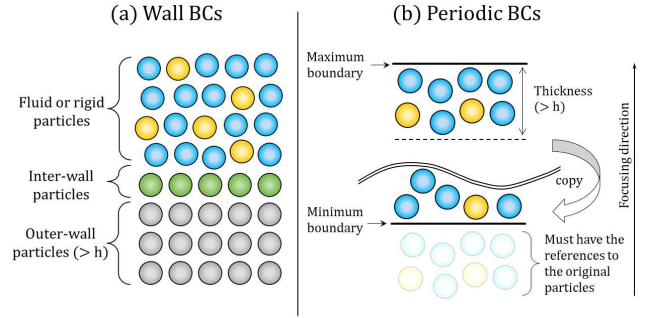}}
\caption{Schematic of boundary conditions.}
\label{fig:CompareBC}
\end{figure}

Let us explain the basic concept of wall boundary conditions (BCs) and periodic BCs. The left part of Fig.~\ref{fig:CompareBC} depicts a schematic of wall BCs. First, we put the frozen particles on the boundary of the simulation domains at a fixed interval, as shown by the green-colored particles (inner-wall particles). Next, we pile up the frozen particles outside the inner-wall particles, as shown by the gray-colored particles (outer-wall particles). We pile up the outer-wall particles so that it is thicker than the size of the kernel radius $h$. Meanwhile, the right part of Fig.~\ref{fig:CompareBC} depicts a schematic of periodic BCs. In periodic BCs, the particles going out from one direction flow in the opposite direction. We copy the particles existing in the area between the maximum boundary and the edge at a depth of the kernel radius $h$ from the boundary to a temporary area adjacent to the opposite boundary of the direction (and vice versa); we call these kinds of particles ``halo particles'' in this paper.

\begin{figure*}[t]
\centerline{\includegraphics[scale=0.43]{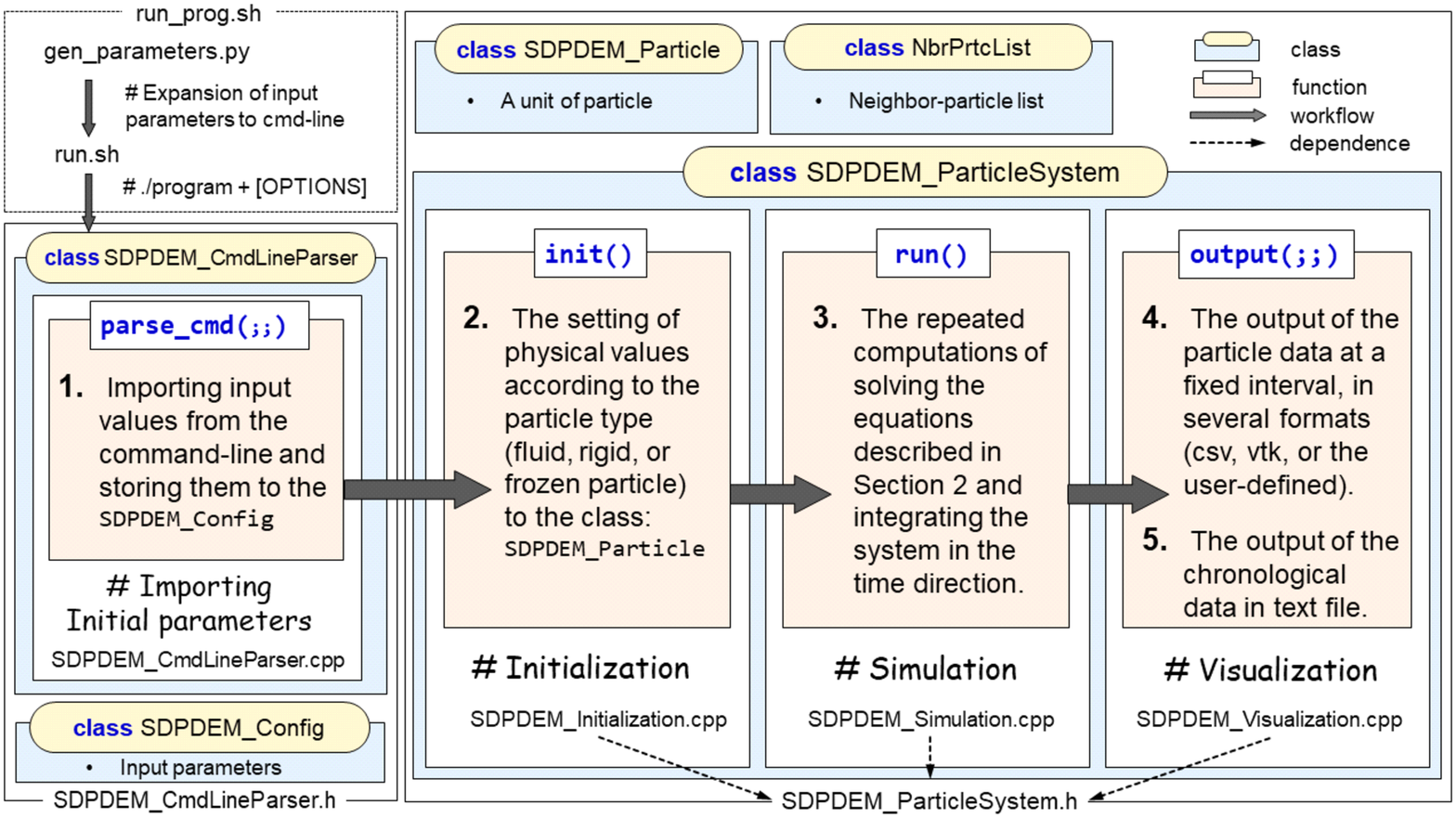}}
\caption{Schematic diagram of our particle simulation system.}
\label{fig:DesignInstr}
\end{figure*}

In simulations, we update the density and pressure of all the kinds of particles other than outer-wall particles,~i.e., fluid, rigid, halo, and inner-wall particles. The positions of outer-wall particles and inner-all particles are fixed during simulations, but the velocities of inner-wall particles are updated the same as fluid or rigid particles. In our code, eight different BCs are selectable: fully wall BCs, fully periodic BCs, and six kinds of wall-periodic hybrid BCs. Besides, non-slip conditions or slip conditions are also selectable. In non-slip conditions, the sign of the velocities of fluid or rigid particles is to be reversed when they approach the edge of the simulation domain. In addition, for the surface treatments, a lack of particles around the interface causes negative pressure; therefore, we correct the negative pressure to be zero if it emerges. 

\subsection{Time integrations}\label{sec:timeinteg}
The total force $\vec{F}_{i}^{\rm total}$ acting on the $i$th particle is computed as the sum of $\vec{F}_{ij}^{\rm SDPD}$, $\vec{F}_{ij}^{\rm DEM}$, $\vec{F}_{ij}^{\rm LUB}$, $\vec{F}_{ij}^{\rm FPI}$, and $\vec{F}_{i}^{\rm EXT}$ according to Eq.~(\ref{eq:govern}). 
At this time, the torque $\vec{N}_{ij}^{\rm total}$ between the $i$th and $j$th particles, is generated by $\vec{F}_{ij}^{\rm SDPD}$ and $\vec{F}_{ij}^{\rm DEM}$.
To summarize, the motion of a particle is described as
\begin{eqnarray}
m_{i}{\frac{d\vec{v}_{i}}{dt}} &=& \vec{F}_{i}^{\rm total}, \label{eq:newton1} \\ 
I_{i}{\frac{d\vec{\omega}_{i}}{dt}} &=& \vec{N}_{i}^{\rm total}. \label{eq:newton2} 
\end{eqnarray}
Here, $m_{i}$, $I_{i}$, $\vec{v}_{i}$, and $\vec{\omega}_{i}$ are the mass, inertia, velocity, and spin angular velocity of the $i$th particle, respectively.
$\vec{N}_{i}^{\rm total}$ is computed by $\sum_{ij} \vec{e}_{i}^{p}\times (\vec{F}_{ij}^{\rm SDPD} + \vec{F}_{ij}^{\rm DEM})$, where $\vec{e}_{i}^{p}$ represents a vector from the $i$th particle to the $j$th particle. The size of $\vec{e}_{i}^{p}$ corresponds to the radius when the particle is rigid and corresponds to half of the initial particle interval when the particle is fluid.
Equation~(\ref{eq:newton1}) and Eq.~(\ref{eq:newton2}) are integrated using the velocity-Verlet algorithm~\cite{10.5555/76990}.
Note that the descriptions of Eq.~(\ref{eq:newton1}) and Eq.~(\ref{eq:newton2}) are slightly different from the original version of SDPD+a in ~\cite{MULLER2015301} in that the mass and radius of the $j$th particle are used in the time integration.
We adopt to use Eq.~(\ref{eq:newton1}) and Eq.~(\ref{eq:newton2}) because these descriptions fit the general view of Newton's second law better.
Note that Eq.~(\ref{eq:newton1}) and Eq.~(\ref{eq:newton2}) are commonly used in the DEM algorithm~\cite{doi:10.1080/19648189.2008.9693050}.

\section{Implementation}
\subsection{Design concept of the simulation system}
Figure~\ref{fig:DesignInstr} shows a schematic diagram of our particle simulation systems. Two classes of \verb|SDPDEM_CmdLineParser| and \verb|SDPDEM_Config| manage the input parameters; the class \verb|SDPDEM_CmdLineParser| parses the command-line parameters and stores them as input parameters in the members of the class \verb|SDPDEM_Config|. 

The procedures of the class \verb|SDPDEM_ParticleSystem| are divided into three parts: initialization, simulation, and visualization. In the initialization, the class creates an array of the particle class \verb|SDPDEM_Particle| and sets initial values of all the particles using \verb|SDPDEM_config|. In the simulation, the system iteratively solves the equations of the SDPD-DEM coupled models and updates them in the time direction. Finally, in the visualization, the system outputs particle data at a fixed interval in several different formats (e.g., csv, vtk, or user-defined).
In addition, the bottom part of Fig.~\ref{fig:DesignInstr} represents the dependancies of major headers and sources of our code. For more detailed information, see the document in the \verb|html/index| in the work directory of our code.

\begin{table*}[t]
\begin{center}
\scalebox{1.0} 
{
\begin{tabular}{lclclc}
\hline
parameters   & notes& parameters& notes & parameters &notes\\ \hline
{\tt --E    }& {\small energy}          &{\tt --scale\_dens}& {\small see \S~\ref{seq:ipparam}} &{\tt --orgx            }&{\small x-origin}\\ 
{\tt --unit }& {\small unit length}     &{\tt --scale\_diam}& {\small see \S~\ref{seq:ipparam}} &{\tt --orgy            }&{\small y-origin}\\
{\tt --mass }& $m_x$                    &{\tt --gx         }& {\small x-gravity}                &{\tt --orgz            }&{\small z-origin}\\
{\tt --iner} & $I_x$                    &{\tt --gy         }& {\small y-gravity}                &{\tt --Coeff\_fcij     }&{\small see \S~\ref{seq:ipparam}} \\
{\tt --xi   }& $\xi$                    &{\tt --gz         }& {\small z-gravity}                &{\tt --Coeff\_fdij     }&{\small see \S~\ref{seq:ipparam}} \\
{\tt --eta  }& $\eta$                   &{\tt --fx         }& {\small x-force }        &{\tt --Coeff\_frij     }&{\small see \S~\ref{seq:ipparam}} \\
{\tt --dens0}& $\rho_0$                 &{\tt --fy         }& {\small y-force }        &{\tt --Coeff\_ftij     }&{\small see \S~\ref{seq:ipparam}} \\
{\tt --pres0}& $p_{0}$                  &{\tt --fz         }& {\small z-force }        &{\tt --itr\_start      }&{\small beginning step} \\
{\tt --beta }& $\beta$                  &{\tt --fillrate\_x}& {\small see \S~\ref{seq:ipparam}} &{\tt --itr\_stop       }&{\small ending step} \\
{\tt --alpha}& $\alpha$                 &{\tt --fillrate\_y}& {\small see \S~\ref{seq:ipparam}} &{\tt --itr\_actv\_rigid}&{\small see \S~\ref{seq:ipparam}}  \\
{\tt --kBT  }& $k_{B}T$                 &{\tt --fillrate\_z}& {\small see \S~\ref{seq:ipparam}} &{\tt --periodic\_type  }&{\small see \S~\ref{seq:ipparam}} \\
{\tt --h    }& $h$                      &{\tt --dt         }& $dt$                              &{\tt --slipcond\_type  }&{\small see \S~\ref{seq:ipparam}} \\
{\tt --dx   }& {\small initial distance}&{\tt --sqrtdt     }& $\sqrt{dt}$                       &{\tt --gravity\_type   }&{\small see \S~\ref{seq:ipparam}} \\
{\tt --cs   }& {\small speed of sound}  &{\tt --vol\_f     }& {\small see \S~\ref{seq:ipparam}} &{\tt --enable\_artvis  }&{\small see \S~\ref{seq:ipparam}} \\
{\tt --Rdem }& $R_x${\small~(radius)}           &{\tt --vol\_s     }& {\small see \S~\ref{seq:ipparam}} &{\tt --enable\_load\_rp}&{\small see \S~\ref{seq:ipparam}} \\
{\tt --Edem }& $E_x${\small~(Young's modulus)}  &{\tt --Lx         }& {\small x-domain size           } &{\tt --N\_intvl\_outvis}&{\small see \S~\ref{seq:ipparam}} \\
{\tt --Pdem }& $\nu_x${\small~(Poisson's rato)} &{\tt --Ly         }& {\small y-domain size           } &{\tt --N\_intvl\_bcalgn}&{\small see \S~\ref{seq:ipparam}} \\
{\tt --Fdem }& $\mu_x${\small~(Friction coeff)} &{\tt --Lz         }& {\small z-domain size           } &{\tt --N\_intvl\_backup}&{\small see \S~\ref{seq:ipparam}} \\ \hline
\end{tabular}
}
\end{center}
\caption{All input parameters (odd-number column) and their explanations~(even-number column).}
\label{table:ipparam}
\end{table*}
\subsection{Setting of initial parameters}\label{seq:ipparam}
In Table.~\ref{table:ipparam}, the odd-number columns show input parameters, and the even-number columns show notes. 
Some parameters are associated with the corresponding notations in Section~2. Other settings are described in the following sections. 

The parameter \verb|--E| is a unit of energy to scale the parameter of energy level of the system $k_B T$. The parameter \verb|--unit| is a unit length to rescale the representative length and is set to one as default. The parameter \verb|--dx| is the distance between particles at the initial state. The parameter \verb|--cs| is the speed of sound. The parameter \verb|--scale_dens| represents the ratio of the density of a rigid particle to that of a fluid particle, and the parameter \verb|--scale_diam| represents the ratio of the diameter of a rigid particle to that of a fluid particle; in this way, the densities and radii of rigid particles are given as relative values of fluid particles in our code. 

A set of (\verb|--gx|, \verb|--gy|, \verb|--gz|) represents the gravities, and (\verb|--fx|, \verb|--fy|, \verb|--fz|) represents the external forces in the respective directions. Note that the difference between the former and latter sets is that the latter one is divided by the mass of a particle. (\verb|--Lx|, \verb|--Ly|, \verb|--Lz|) designates the size of the simulation domain. (\verb|fillrate_x|, \verb|fillrate_y|, \verb|fillrate_z|) indicates the filling ratio of fluid in each direction.

Continuously, \verb|--vol_f| is the spherical volume with the radius of $h$. \verb|--vol_s| is the spherical volume of a rigid particle. (\verb|--orgx|, \verb|--orgy|, \verb|--orgz|) indicates the vector coordinates of the origin, which correspond to the minimum coordinates of the simulation domain. Each parameter from \verb|--Coeff_fcij| to \verb|--Coeff_ftij| is the coefficient of the four of SDPD forces; users can switch the effects of these four forces on or off in Eq.~(\ref{eq:FijC}) to Eq.~(\ref{eq:FijT}).

The reminding parameters mostly represent the computational parameters; \verb|--itr_start| and \verb|--itr_stop| indicate the beginning and end steps of a simulation. The parameter \verb|--actv_rigid| indicates the beginning step of rigid particles starting to be updated and interact with other particles. \verb|--actv_rigid| is set to zero as default. 

Also, \verb|--periodic_type| indicates the boundary conditions~(wall BCs, periodic BCs, or six types of wall-periodic hybrid BCs). 
\verb|--slipcond_type| designates the slip conditions or non-slip conditions. 
\verb|--gravity_type| indicates the type of gravity,~i.e., gravity in one direction or that in the opposing direction, the latter of which is used in the reverse-Poiseuille simulations~(see Section~\ref{sec:casestdy}). 

\verb|--enable_artvis| designates whether we set the artificial viscosity~\cite{SHAKIBAEINIA201213} or not. 
\verb|--N_intvl_outvis| represents the interval of the output of visualization files. Besides, \verb|--N_intvl_backup| indicates the output interval of restart files; we can always stop and restart the simulations using the backed-up files. 
In the end, \verb|--N_intvl_pcalgn| designates the interval of putting rigid particles in a simulation domain at the initial state; e.g., when we set {\verb|--N_intvl_pcalgn| to be five,~i.e., one rigid particle after arranging every four fluid particles.

If we set the parameter \verb|--N_intvl_pcalgn| to $\rm -1$, there are no rigid particles arranged.~i.e., only fluid particles are put in the simulation domain. In such a case, users can import the relative positions of rigid particles from the list in the supplementary file \verb|input_rigid_particle.csv| by setting the parameter \verb|--enable_load_rp| to \verb|True|. In this case, the fluid particle located closest to the imported particle is replaced by the imported particle; e.g., when we designate the relative coordinates as $\rm (0.5,~0.85,~0.5)$, the location of the particle is interpreted as $\rm (0.5Lx+min.x, 0.85Ly+min.y, 0.5Lz+min.z)$. 

Note that users can check whether the input parameters are appropriately set to the class \verb|SDPDEM_Config| using the log file \verb|list_of_input_parameters.txt|.

\subsection{Data structure of particles}\label{seq:datastrcpc}
All the particle data are recorded in a sequential array of a unified particle class that has both variables of the DEM and SDPD. We create an array of the class \verb|SDPDEM_Particle|, which has the following data structure:

\begin{lstlisting}[mathescape=true, numbers=left, label=pcstrct, frame=lines, caption= Structure of an SDPDEM particle.]
Class SDPDEM_Particle {
  // (1) Physical variables
  Real mass; // Mass
  Real iner; // Inertia
  Real dens; // Density
  Real pres; // Pressure
  Real3 pos; // Position
  Real3 vel; // Velocity
  Real3 angvel; // Angular velocity
  Real3 F_fpi_DG; // Drag force in $\rm \color[rgb]{0,0.5,0}{\tt Eq.(\ref{eq:F_fpi_DG})}$
  Real3 F_fpi_PG; // $\rm \color[rgb]{0,0.5,0}{\tt \Delta P}$ force in $\rm \color[rgb]{0,0.5,0}{\tt Eq.(\ref{eq:F_fpi_PG})}$
  Real3 F_dem; // Contacting force in$~\rm \color[rgb]{0,0.5,0}{\tt \S.\ref{sec:rigidforce}}$
  Real3 F_sdpd; // Fluid force in$~\rm \color[rgb]{0,0.5,0}{\tt \S.\ref{seq:fluidforce}}$
  Real3 F_total; // Total force in $~\rm \color[rgb]{0,0.5,0}{\tt \S.\ref{sec:timeinteg}}$
  Real3 N_dem; // Torque by contacting force in$~\rm \color[rgb]{0,0.5,0}{\tt \S.\ref{sec:timeinteg}}$
  Real3 N_sdpd; // Torque by fluid force in$~\rm \color[rgb]{0,0.5,0}{\tt \S.\ref{sec:timeinteg}}$
  Real3 N_total; // Total torque in $~\rm \color[rgb]{0,0.5,0}{\tt \S.\ref{sec:timeinteg}}$
  Real Rdem; // Radius in$~\rm \color[rgb]{0,0.5,0}{\tt \S.\ref{sec:rigidforce}}$
  Real Edem; // Young's modulus in$~\rm \color[rgb]{0,0.5,0}{\tt \S.\ref{sec:rigidforce}}$
  Real Pdem; // Poisson's ratio in$~\rm \color[rgb]{0,0.5,0}{\tt \S.\ref{sec:rigidforce}}$
  Real Ddem; // $\color[rgb]{0,0.5,0}{\tt D}$ parameter of $\rm \color[rgb]{0,0.5,0}{\tt Eq.(\ref{eq:deminputC})}~$in$~\rm \color[rgb]{0,0.5,0}{\tt \S.\ref{sec:rigidforce}}$
  Real fvf; // Fluid volume fraction in $~\rm \color[rgb]{0,0.5,0}{\tt \S.\ref{sec:fpiforce}}$
  AoS<NbrPrtcList> dSij; // Neighbor-particle list
 	
  // (2) Computational variables
  int id_grid; // ID of a grid in background grids 
  int id_grid_x, id_grid_y, id_grid_z; // Ditto 
  int id_link; // ID of a particle in the same chain 
  int id_orgp; // ID of the copy-source particle
  int id_type; // Type of this particle 
  int id_global; // Serial ID of this particle
}
\end{lstlisting}

Lines~3-9 describe the fundamental properties: mass, inertia, density, pressure, position, velocity, and angular velocity (the class has the same set of these seven variables for temporary use, but they are omitted in List.~\ref{pcstrct} for ease of visualization). Each of lines~10-14 describes the summed up force of the corresponding pair-wise force explained in Section~2 with regard to the neighboring particles. Lines~15-17 describe the torques. Meanwhile, lines~18-21 show the parameters that specify the rigidity of the particle. Line~22 expresses the fluid volume fraction of the particle ~$\epsilon$, which is calculated as 
\begin{eqnarray}
\epsilon~=~|V_{f}-V_{s}N_{r}|/V_{f}, \label{eq:fvfdef} 
\end{eqnarray}
where $V_{f}$ is the spherical volume of radius $h$, and $V_{s}$ is the spherical volume of a rigid particle. $N_{r}$ represents the number of rigid particles in the neighboring area. 

{\tt dSij} is an array of the class \verb|NbrPrtcList|, which has the following data structure:
\begin{lstlisting}[mathescape=true, numbers=left, label=pcstrctsub, frame=lines, caption= Structure of~a neighbor-particle list.]
Class NbrPrtcList {
  // (1) Physical variables
  /* Independent Wiener increments */
  Matrix3d dWM_ij; // $\tt \color[rgb]{0,0.5,0}{\tt d\vec{W}_{ij}}$ in $\rm \color[rgb]{0,0.5,0}{\tt Eq.(\ref{eq:reldWij})}$
  /* Tangential displacement in DEM algorithm */ 
  Real3 disp_tij; // $\color[rgb]{0,0.5,0}{\tt \vec{\delta}_{{t}_{ij}}^{n}}$ in $\rm \color[rgb]{0,0.5,0}{\tt Eq.(\ref{eq:tandispn})}$

  // (2) Computational variables	
  int index; // ID of this particle 
  int id_type; // Type of this particle 
  bool flag_contact; // Contacting or non-contacting
}
\end{lstlisting}
Here, \verb|dWM_ij| is an independent Wiener increment ${\tt d\vec{W}_{ij}}$ in Eq.(\ref{eq:reldWij}), and 
\verb|disp_tij| is the tangential displacement ${\tt \vec{\delta}_{{t}_{ij}}^{n}}$ in Eq.(\ref{eq:tandispn}). 
The remaining parameters are explained later.

In our implementation, we introduce a technique of neighbor-particle lists as follows. 
First, we cover the simulation domain by spatial background grids. We then register one of the IDs of the particles existing in each grid to the grid. At this time, constructing a linked-list structure among the particles belonging to the same grid makes it possible to reduce the memory size of the background grids because only a single particle existing in each grid is registered to the grid~\cite{GREST1989269}. 

Subsequently, each particle searches its neighbors by tracing the linked-lists and creating an array of \verb|NbrPrtcList| with the size of the amount of neighboring particles.
This local list creation is performed at the beginning of each computational step; each particle refers to its own local list in the remaining parts.
Figure~\ref{fig:NbrPrtcList} shows a schematic of two examples of local lists that the particles have in themselves.
\begin{figure}[h]
\centerline{\includegraphics[scale=0.25]{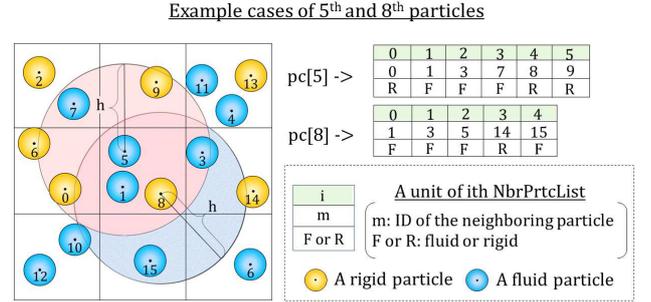}}
\caption{Schematic of the neighbor-particle list.}
\label{fig:NbrPrtcList}
\end{figure}

In List~\ref{pcstrct}, {\tt id\_grid}, {\tt id\_grid\_k (k=x,y,z)} are the parameters that indicate the index of a grid in which the particle exists. {\tt id\_link} indicates one of the particles belonging to the same chain of linked-lists. {\tt id\_global} is a serial index assigned to all the particles. The parameter {\tt id\_orgp} is valid when the particle is a halo particle; {\tt id\_orgp} indicates an index of the original particle, of the halo particle. The {\tt id\_type} represents the type of particle: the fluid, rigid, wall, or halo.

Meanwhile, in List~\ref{pcstrctsub}, {\tt index} represents the index of the neighboring particle. {\tt id\_type} represents the type of neighboring particle.
In the end, the parameter {\tt flag\_contact} specifies that the neighboring particle is in either a state of contacting (true) or non-contacting (false).
{\tt flag\_contact} is determined by measuring the distance between the host particle and the neighboring particle in the list. The particle is eliminated from the neighbor-particle list if {\tt flag\_contact} is false. If {\tt flag\_contact} is true but the particle does not exist on the list, the particle is newly added to the neighbor-particle list.

\subsection{Overviews of computational proceses}
In our code, the main loop function {\tt run()} calls a top subroutine of \verb|compute_total_processes()|, which computes the mechanics of the SDPD-DEM coupled system. 
Algorithm~\ref{alg:compflow} shows pseudo code of \verb|compute_total_processes()|.
\begin{algorithm}[h]
\caption{\tt compute\_total\_processes()} \label{alg:compflow}
\begin{algorithmic}[1]
\For{${\rm step} = {\tt itr\_start} \to {\tt itr\_stop}$}
\For{$\rm velver = {\tt 1} \to {\tt 2}$}
\State {\tt compute\_neighbor\_list(velver);} 
\State {\tt compute\_particle\_dens(step);} 
\State {\tt compute\_particle\_pres();} 
\State {\tt smoothe\_particle\_pres();} 
\State {\tt compute\_force();} 
\State {\tt update(velver, step);} 
\State {\tt output();} 
\EndFor
\EndFor
\end{algorithmic}
\end{algorithm}

Line~1 describes the iteration step from \verb|--itr_start| to \verb|--itr_stop|. Line~2 shows the subloop of the velocity-Verlet method.
In \verb|compute_neighbor_list()| in line~3, the system creates the background grids and linked-list structures in each grid. After that, each particle creates an array of the class \verb|NbrPrtcList| by tracing the linked lists in the aforementioned manner in Section~\ref{seq:datastrcpc}. The \verb|index|, \verb|id_type|, and \verb|flag_contact| in \verb|NbrPrtcList| are set depending on the types of neighboring particles. Besides, \verb|disp_tij| is computed in the process of the DEM. 
Because of the symmetricity of independent Wiener increments $\rm d\vec{W}\_{ij}$, we set the same random matrix of \verb|dMW_ij| in each pair of neighboring particles. 
Each element of \verb|dMW_ij| is computed using the library \verb|MT19937-64|~\cite{10.1145/369534.369540, 10.1145/272991.272995}.
When either of the particles is a halo particle, we set the same \verb|dMW_ij| to the original particle designated by \verb|id_orgp| of the halo particle.

In \verb|compute_particle_dens()| in line~4, each particle computes the density using Eq.~(\ref{eq:compdens}). To suppress the discretization error, normalization of the coefficient of the kernel function and the reinitialization of the density~\cite{COLAGROSSI2003448} are introduced in our code. In \verb|compute_particle_pres()| in line~5, each particle computes the pressure using Eq.~(\ref{eq:prtcpres}).
Additionally, in \verb|smoothe_particle_pres()| in line~6, each particle smoothes the pressure by a weighted calculation~\cite{Imoto2019} in a similar manner to Eq.~(\ref{eq:compdens}).

\verb|compute_force()| in line~7 is composed of five subroutines, as shown in Algorithm~\ref{alg:compforce}. In \verb|compute_fluid()| in line~1, each fluid particle computes the fluid force $\vec{F}_{ij}^{\rm SDPD}$ among all the neighboring particles including rigid particles, on the basis of the single pressure hypothesis. Subsequently, each rigid particle computes the pressure gradient force $\vec{F}_{ij}^{\rm PG}$ from each of the neighboring fluid particles using Eq.~(\ref{eq:F_fpi_PG}).
\begin{algorithm}[h]
\caption{\tt Breakdown of compute\_force()} \label{alg:compforce}
\begin{algorithmic}[1]
\State {\tt compute\_fluid();} 
\State {\tt compute\_fpi\_action();} 
\State {\tt compute\_fpi\_reaction();} 
\State {\tt compute\_rigid();} 
\end{algorithmic}
\end{algorithm}

In \verb|compute_fpi_action()|, each fluid particle calculates the fluid volume fraction $\epsilon$ using Eq.~(\ref{eq:fvfdef}) and then computes $\vec{F}_{ij}^{DG}$ using Eq.~(\ref{eq:F_fpi_DG}) and Eq.~(\ref{eq:beta}). Subsequently, in \verb|compute_fpi_reaction()|, each rigid particle computes the reaction force from the neighboring fluid particles using Eq.~(\ref{eq:F_fpi_DG_reac}). 
In \verb|compute_rigid()|, each rigid particle computes the contacting forces among rigid particles using the DEM from Eq.~(\ref{eq:dem_fnij}) to Eq.~(\ref{eq:lubri}). 
After computing the total force and torque, in \verb|update(velver, step)| in Algorithm~\ref{alg:compflow}, the system is integrated by the velocity-Verlet method. In the end, in \verb|output()|, the system outputs particle data at a fixed interval of the computational iterative loop.
\begin{lstlisting}[mathescape=true, numbers=left, label=maincpp, frame=lines, caption= main.cpp.]
#include ``SDPDEM_ParticleSystem.h''
int main (int argc, char* argv[])
{
	SDPDEM_ParticleSystem ps (argc, argv, ``$\color[rgb]{0.8,0,0}{\tt Result}$'', ``$\color[rgb]{0.8,0,0}{\tt file\_}$'');
    ps.run(); 
    return 0;
}
\end{lstlisting}

List~\ref{maincpp} shows the \verb|main.cpp| file of our source code. The third argument in the declaration of class \verb|SDPDEM_ParticleSystem| in line 4 indicates the name of the directory in which all the resulting data are output. The fourth argument indicates the prefix name of sequentially numbered files. In our default setting, the back-up files to stop or restart simulations are exported to the directory \verb|Result/backup/| at the interval designated by \verb|--N_intvl_backup|. 
Similarly, the system outputs the visualization files to the directory of \verb|Result/vtp/| at the interval indicated by \verb|--N_intvl_outvis|. The position, speed, density, pressure, and two kinds of particle types~(\verb|pc_type_seperated|, or \verb|pc_type_merged|) are exported to the sequentially numbered files in a particle VTK format. \verb|pc_type_seperated| sets the different IDs to the particles according to their types (fluid, rigid, inner-wall, or outer-wall). In contrast, \verb|pc_type_merged| assigns $1$ to the fluid or rigid particles and $2$ to other types of particles. 

\begin{figure}[h]
\centerline{\includegraphics[scale=0.25]{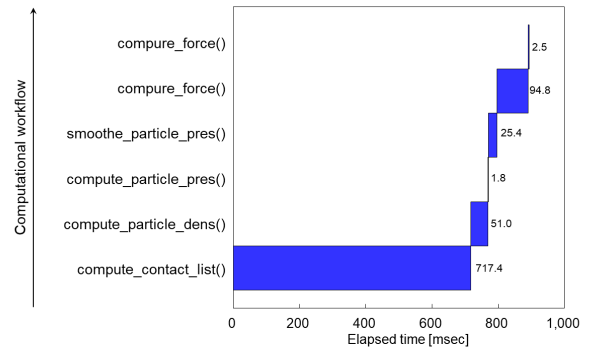}}
\caption{A breakdown of the total computational time.}
\label{fig:CompTimeTotal}
\end{figure}

Figure~\ref{fig:CompTimeTotal} shows a comparison of the calculation times consumed for the respective modules in Algorihms~\ref{alg:compflow} and \ref{alg:compforce}. Because our system performs the neighbor-particle search on the background grids only one time,  the calculation time for \verb|compute_neighbor_list()| is dominant in the total calculation time. Conversely, the calculation time for each module in \verb|compute_force()| can be said to be reduced due to avoiding repeated neighbor-particle searches by introducing the local neighbor-particle lists of particles.

\section{Case studies}\label{sec:casestdy}
\subsection{Reverse-Poiseuille flow}
Following the literature, we validate our code by demonstrating simulations of reverse-Poiseuille flow~\cite{MULLER2015301}, which is a common benchmark for measuring the viscosity of the particle fluid model~\cite{doi:10.1063/1.1883163, LEI2017571}.
In a rectangular-shaped domain, the flow is divided into the upper and lower parts in the vertical direction, and each flow is driven by the same sized force in the opposite direction under periodic BCs. The velocity at the boundaries of the simulation domain is stochastically offset and become zero. Thus, we can compare the simulation with the theoretical solution of Hagen-Poiseuille flow~\cite{doi:10.1146/annurev.fl.25.010193.000245}.

In general, the velocity profile of the flow driven in the y-direction in between the two plates, which is perpendicular to z-direction, is given by~\cite{doi:10.1063/1.1883163} 
\begin{eqnarray}  
v(z) = \frac{\rho g_y}{2\mu_{f}}(zD-z^2). \label{eq:poiseuanasol}
\end{eqnarray}
Here, $v(z)$ and $g_{y}$ represent the speed and gravity in the y-direction, respectively, $D$ indicates the distance between two plates, $\mu_{f}$ represents the viscosity of fluid, and $\rho$ indicates the density of fluid.
By integrating Eq.~(\ref{eq:poiseuanasol}) with respects to $z$ between $0$ and $D$, we obtain the averaged velocity as 
\begin{eqnarray}  
V_{\rm rp} = \frac{\rho g_{y}D^2}{12\mu_{f}}. \label{eq:poiseuave}
\end{eqnarray}
The viscosity of the reverse-Poiseuille flow is obtained by simultaneously fitting Eq.~(\ref{eq:poiseuave}) to the convex and concave curves of the flow.

In this test, all the input parameters are given the same values as used in the simulation test reported in~\cite{MULLER2015301}; the simulation domain of $\rm (Lx, Ly, Lz)$ is set to $\rm (20, 40, 10)$, and each unit of $\rm (E, unit, mass)$ is set to one. We set the parameter $\eta$ to $25$ and set the parameter $\xi$ so that it satisfies the relationship of $\xi=20\eta/21$. The density of $\rho_0$ is set to $\rm 3mass/unit^3$. Both values of the pressure $p_{0}$ and the parameter $\beta$ are set to $\rm 100E/unit^3$. We set the gravitiy to zero in each direction assuming mesoscale flow. Instead, we drive the particles by the pressure-gradient force of $\rm (fx, fy, fz)$, which is set to $(0, 1.5, 0)$ in this simulation test. Two types of BCs are compared in the simulations; Figure~\ref{fig:CompRevPoiseuilleBC}(a) shows fully-periodic BCs, and Fig.~\ref{fig:CompRevPoiseuilleBC}(b) shows wall-periodic hybrid BCs, where the wall BCs and periodic BCs are respectively given in the z-direction and xy-plane direction. 
For more details, refer to the attached file~\verb|gen_parameters|\verb|_revpoiseuille00.py|.

In Fig.~\ref{fig:CompareWallPeriRev}, the triangle symbols represent the measured points in the case of (a) fully-periodic BCs. The blue-colored solid curve is obtained by fitting Eq.~(\ref{eq:poiseuave}) to the measured points of fully-periodic BCs using the least-squares method. In contrast, the circle symbols represent the measured points when using the wall-periodic hybrid BCs. The red-colored dashed curve is obtained by fitting Eq.~(\ref{eq:poiseuave}) to the measured points of the wall-periodic hybrid BCs in a similar manner. The relative error of the viscosity using (b) to that using (a) was at most $2$ percent in our test. Discussion on the difference between (a) and (b) when simulating reverse-Poiseuille flow to measure the viscosity is a point of emphasis in this paper; the results in Fig.~\ref{fig:CompareWallPeriRev} suggest that the use of wall BCs is permissible within the level of accuracy used.

\begin{figure}[t]
\centerline{\includegraphics[scale=0.25]{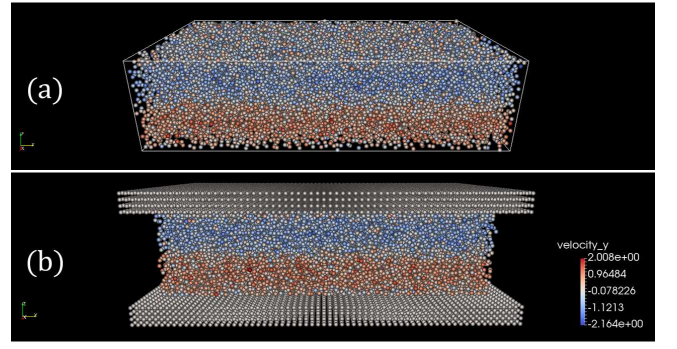}}
\caption{Simulations of the reverse-Poiseuille flow with (a) fully-periodic BCs and (b) wall-periodic hybrid BCs.}
\label{fig:CompRevPoiseuilleBC}
\end{figure}

\begin{figure}[t]
\centerline{\includegraphics[scale=0.30]{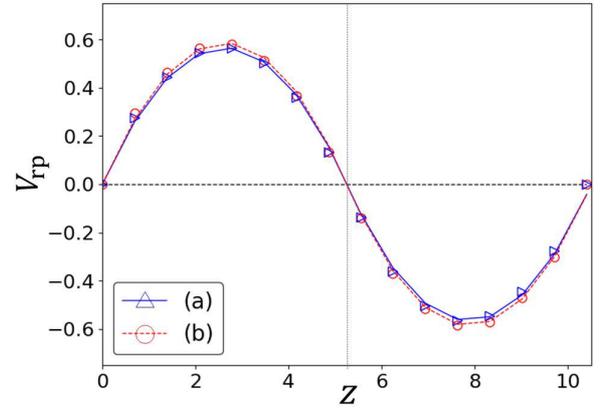}}
\caption{Comparisons of the velocity profile between (a) fully-periodic BCs and (b) wall-periodic hybrid BCs.}
\label{fig:CompareWallPeriRev}
\end{figure}

\subsection{Single-particle sedimentation} \label{sec:prtcsediment}
We performed a simulation of particle sedimentation to validate our code of the SPH-DEM coupled model. A single particle is fixed in a fluid, as shown in Fig.~\ref{fig:SetUpPrtcSediment}. 
\begin{figure}[t]
\centerline{\includegraphics[scale=0.25]{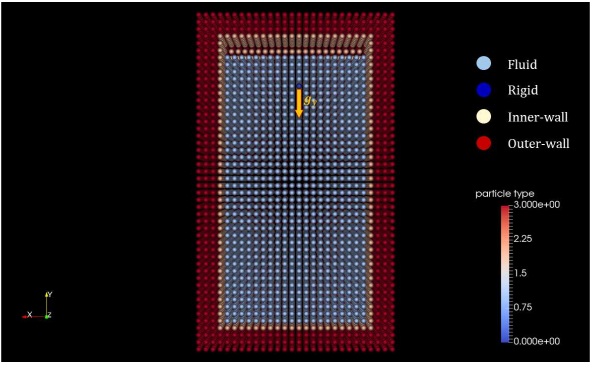}}
\caption{Set up of the single-particle sedimentation problem.}
\label{fig:SetUpPrtcSediment}
\end{figure}
The particle starts to fall after starting the simulation. In this problem, the terminal velocity of the particle is described as follows by considering equilibrium between gravity, buoyancy force, and drag forces~\cite{10.2307/41254479}:
\begin{eqnarray}
V_{\rm tm} = \begin{cases}
\frac{d^2(\rho_{s}-\rho_{f})g}{18\mu_{f}} & ({\rm Re < 2}), \\
\Biggl\{ \frac{4}{225}{\frac{{(\rho_{s}-\rho_{f})}^{2}g^{2}}{\rho_f\mu_{f}}}\Biggr\}^{\frac{1}{3}}d & ({\rm 2 ~ <~Re~< ~500}), \\
\Biggl\{ \frac{4}{3\times 0.44}\frac{(\rho_{s}-\rho_{f})g}{\rho_{f}}d\Biggr\}^{\frac{1}{2}} & ({\rm 500~<~Re}).
	   \end{cases}
	   \end{eqnarray}
Here, the parameter $d$ is the diameter of a rigid particle, $\rho_{f}$ is the density of the fluid, and $\rho_{s}$ is the density of a rigid particle. $\mu_{f}$ is the viscosity of the fluid, and $g$ is gravity in the vertical direction. The regime of $\rm Re < 2$ is called the Stokes regime, that of $\rm 2 < Re < 500$ is the Allen regime, and that of $\rm 500 < Re$ is the Newton regime~\cite{PhysRevE.97.032611}. 

We set the reference density $\rho_{0}$ to ${\rm 1000~kg}$. Let the reference viscosity $\eta$ be $0.01$ ${\rm Pa\cdot s}$, which is approximately ten times as large as the viscosity of water at room temperature. Meanwhile, we set the domain size ${\rm (Lx, Ly, Lz)}$ to be ${\rm (2~m, 4~m, 1~m)}$. Because of Reynolds' law of similarity, this setup corresponds to the case of ${\rm (0.2~m, 0.4~m, 0.1~m)}$ using water. The relative density ratio of rigid particles to fluid is set to ${\rm 2.7}$, assuming the case of aluminum microparticles. 
Particles are put inside the simulation domain with an interval of $\rm dx=0.1$. The number of fluid particles becomes ${\rm 7,599}$ and that of frozen particles becomes ${\rm 16,192}$ in total.
For more detailed parameter settings, see~\verb|gen_parameters_particle_sediment.py|.
\begin{figure}[b]
\centerline{\includegraphics[scale=0.3]{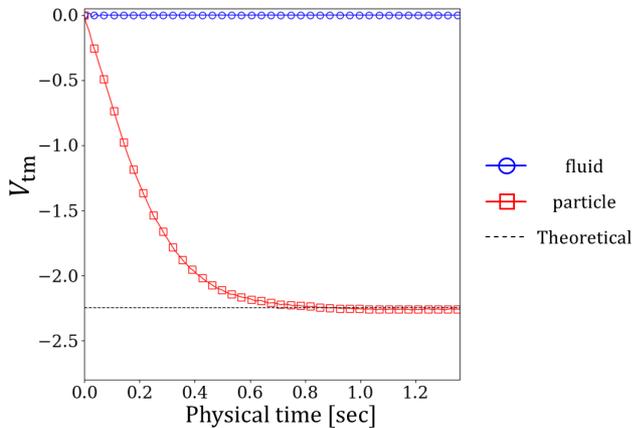}}
\caption{Terminal velocity of the single-particle sedimentation simulation.}
\label{fig:ResultPrtcSediment}
\end{figure}

Figure~\ref{fig:ResultPrtcSediment} indicates the terminal velocity of a single-particle sedimentation simulation. The terminal velocity was confirmed to show a good agreement with the theoretical solution. Note that the Reynolds number was $\rm 2.3\times 10^{4}$, which is in the Newton regime. It is significant that only less than $24,000$ particles are used in this simulation. 

\begin{figure}[b]
\centerline{\includegraphics[scale=0.25]{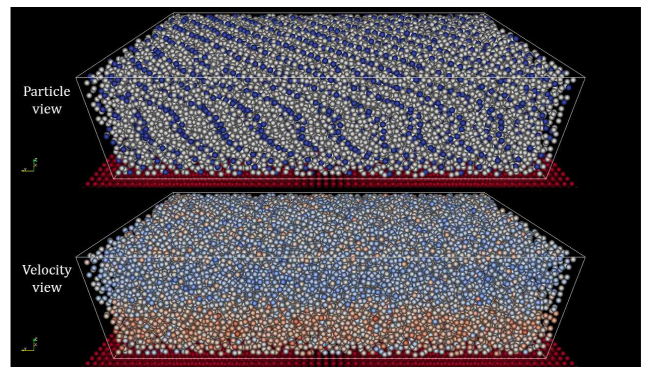}}
\caption{A snapshot of the slurry simulation with a concentration of $11.0~\%$ after elapsing $\rm 7~s$ in physical time.}
\label{fig:ImageRevPoiseulleSlurry}
\end{figure}
\subsection{Reverse-Poiseuille flow of slurry}
We performed the simulations of reverse-Poiseuille flow with several different values of concentration, which is defined as the volume ratio of rigid particles to the total volume. The properties of fluid particles are the same as those of the simulation in Fig.~\ref{fig:CompRevPoiseuilleBC}~(b). At the initial condition, we arrange the rigid particles at the fixed interval, which is designated by the input parameter \verb|N_intvl_pcalgn|. We set the parameter~\verb|N_intvl_pcalgn| to $5$, $10$, and $15$, which correspond to the cases of $3.67$, $5.5$, and $11.0$ percent concentration, respectively. 

Figure~\ref{fig:ImageRevPoiseulleSlurry} shows a snapshot of the slurry simulation with a concentration of $\rm 11\%$ after elapsing $7$ s in physical time from two views: (a) visualization by particle type (fluid, rigid, and frozen particles) and (b) that by velocity profile in the y-direction. The upper wall particles are cut out for ease of visualization. Figure~\ref{fig:CompVelProWallPeriRev} represents comparisons of the velocity profiles with the different concentrations between $0$, $3.67$, $5.5$, and $\rm 11.0\%$. It was confirmed that the amplitude of the velocity deteriorates as the concentration increases. This result is reasonable because the specific gravity of rigid particles is set to $2.7$ times larger than the fluid, as aforementioned. Figure~\ref{fig:ResultRevPoiseulleSlurry} shows the same result from the view of viscosity; vertical lines indicate the relative viscosity, which is obtained by Eq.~(\ref{eq:poiseuave}) using the measured velocity shown in Fig.~\ref{fig:ImageRevPoiseulleSlurry} and scaled by reference viscosity. The relative viscosity was confirmed to increase as the concentration increases. 

\begin{figure}[t]
\centerline{\includegraphics[scale=0.3]{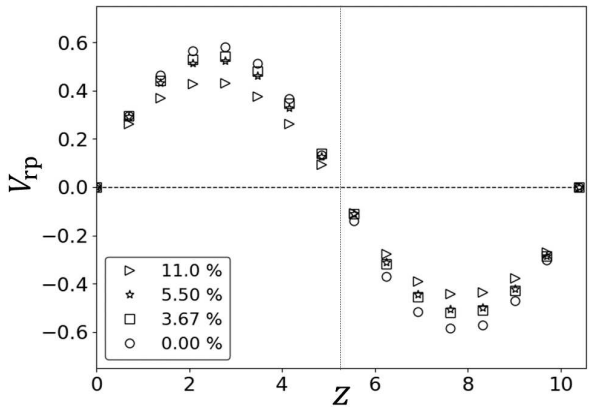}}
\caption{Comparison of the velocity profiles with different concentrations between $0$, $3.67$, $5.5$, and $11.0$ percent.}
\label{fig:CompVelProWallPeriRev}
\end{figure}

\begin{figure}[t]
\centerline{\includegraphics[scale=0.24]{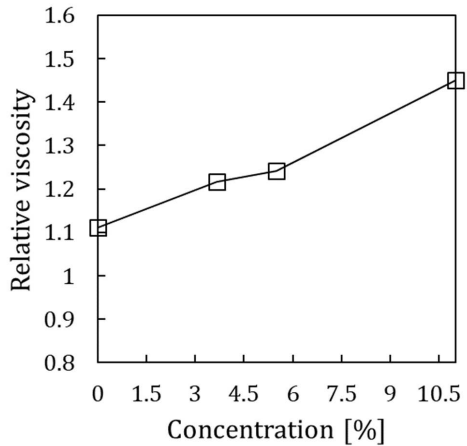}}
\caption{Comparison of the rerative viscosities with different concentrations between $0$, $3.67$, $5.5$, and $11.0$ percent.}
\label{fig:ResultRevPoiseulleSlurry}
\end{figure}

\begin{figure*}[t]
\centerline{\includegraphics[scale=1.0]{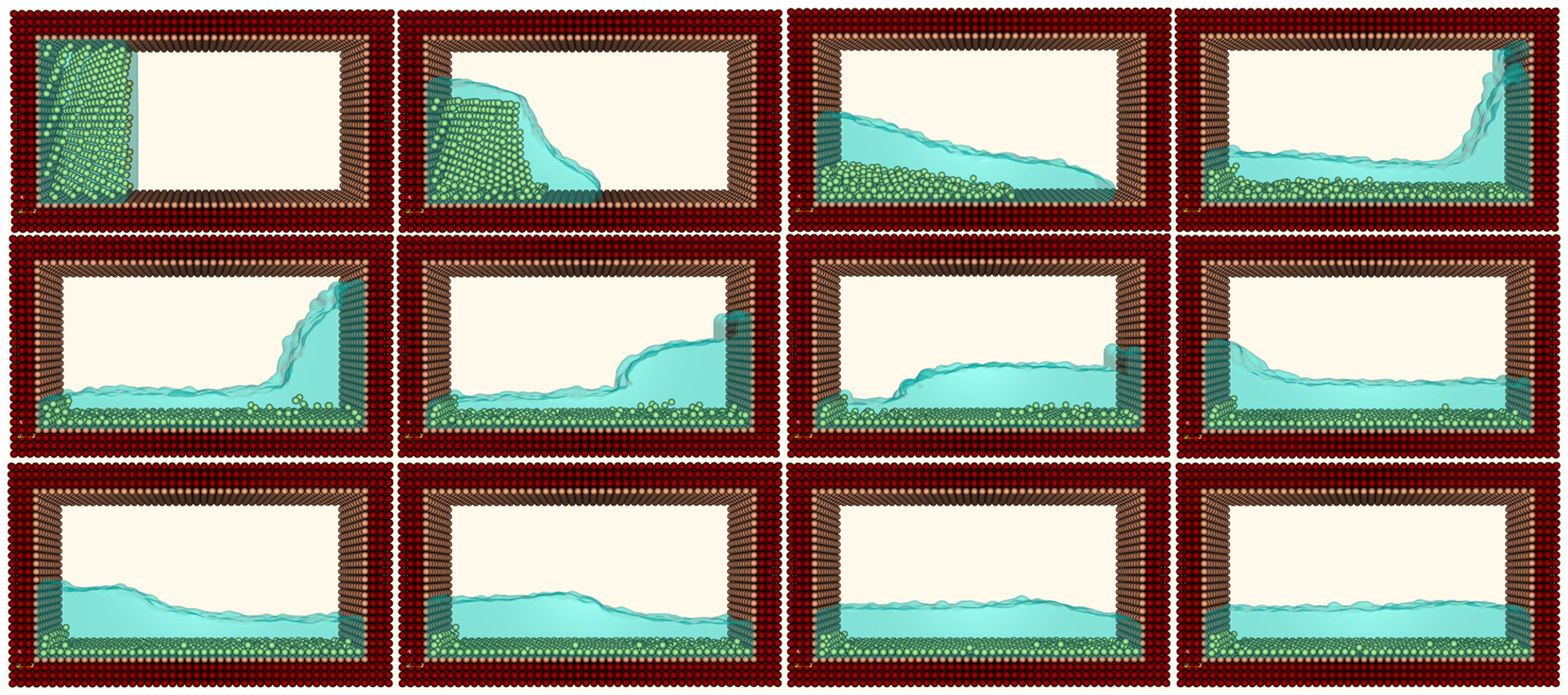}}
\caption{Snapshots of a dam-breaking simulation between $\rm 0$ and $\rm 2.9~s$ in physical time.}
\label{fig:DamBreakingProblem}
\end{figure*}
\subsection{Extension to the related problems}
Because SDPD includes the algorithm of SPH, users can utilize our simulation code for other kinds of problems in related fields. Figure~\ref{fig:DamBreakingProblem} shows a snapshots of a dam-breaking simulation between $\rm 0$ and $\rm 2.9~s$ in physical time.
The number of fluid particles becomes $\rm 1,600$, that of rigid particles becomes $\rm 400$, and that of frozen particles becomes $\rm 16,192$ in total.
It was confirmed that the water pole separates into two phases because of the difference in specific gravity. 
In such free-surface problems, it is relatively difficult not to use artificial viscosity to stabilize the simulations. However, using the artificial viscosity is not appropriate when applying our code to reverse-Poiseuille flows to measure the viscosity. Thus, in our simulation code, we introduce the collision method~\cite{SHAKIBAEINIA201213} to switch it on/off with the parameter \verb|--enable_artvis|, which is set to be true when using the artificial viscosity. Note that the parameter \verb|--enable_artvis| is set to be true in case of the particle sedimentation test in Section~\ref{sec:prtcsediment} and this dam-braking problem.

\section{Conclusion}
In this paper, we presented our new open-source code that works as a viscometer of particle-based simulations of three-dimensional fluid-particle interaction systems, targetting slurry or suspension flow in chemical engineering. SDPD, which is a fluid particle model developed for thermodynamic flow at mesoscale, was appropriately combined with the contact model of the DEM. The mechanics of fluid-particle interaction was modeled by a two-way interaction scheme using a drag-force model. We demonstrated our simulation code through several validation tests: simulations of the reverse-Poiseuille flow, single-particle sedimentation, and a dam-breaking problem containing rigid particles in reasonable calculation time. 

In the simulation of reverse-Poiseuille flow, it was found that wall boundary conditions show a certain agreement with periodic boundary conditions. It was clarified that the use of wall boundary conditions is permissible within this level accuracy in the accurate measurement of viscosity of the fluid. On the other hand, the simulation results of single-particle sedimentation showed good agreement with the theoretical solution. 

In the simulation of the reverse-Poiseuille flows of slurry, a fluid containing rigid particles, we obtained intuitively understandable results that the viscosity increases as the number of heavy rigid particles increases in mesoscale flows. We understand that the current model (an alternative model using SPH-DEM) is still in a development stage when applying it to SDPD-DEM simulations. Nevertheless, it is meaningful that the SPH-DEM model was found to work as an alternative to SDPD-DEM at a certain level. Further studies on the SDPD-DEM coupling model are expected in future work.

We hope our new open-source code is beneficial for scientists, researchers, and engineers in a broader area of physics and that it promotes further studies on SDPD-DEM simulations.

\section*{Acknowledgement}
This research was supported by JSPS KAKENHI Grant Number 18H06459.
The author would like to thank Editage (www.editage.jp) for English language editing.
The author would also like to express my gratitude to my family for their moral support and warm encouragement.





\bibliographystyle{h-physrev3}
\bibliography{reference}

\begin{thebibliography}{10}

\bibitem{JIANG2016949}
J.~Jiang, Z.~Lu, Y.~Niu, J.~Li, and Y.~Zhang,
\newblock Study on the preparation and properties of high-porosity foamed
  concretes based on ordinary portland cement, Materials \& Design {\bf 92},
  949  (2016).

\bibitem{Shkundin1974}
B.~M. Shkundin,
\newblock Clogging of slurry pipelines, Hydrotechnical Construction {\bf 8},
  473 (1974).

\bibitem{PhysRevE.67.026705}
P.~Espa\~nol and M.~Revenga,
\newblock Smoothed dissipative particle dynamics, Phys. Rev. E {\bf 67}, 026705
  (2003).

\bibitem{Espaol1995StatisticalMO}
P.~Espa{\~n}ol and P.~B. Warren,
\newblock Statistical mechanics of dissipative particle dynamics.,
\newblock 1995.

\bibitem{gingold1977smoothed}
R.~A. Gingold and J.~J. Monaghan,
\newblock Smoothed particle hydrodynamics: theory and application to
  non-spherical stars, Monthly notices of the royal astronomical society {\bf
  181}, 375 (1977).

\bibitem{monaghan1992smoothed}
J.~J. Monaghan,
\newblock Smoothed particle hydrodynamics, Annual review of astronomy and
  astrophysics {\bf 30}, 543 (1992).

\bibitem{MULLER2015301}
K.~M{\"{u}}ller, D.~A. Fedosov, and G.~Gompper,
\newblock Smoothed dissipative particle dynamics with angular momentum
  conservation, Journal of Computational Physics {\bf 281}, 301  (2015).

\bibitem{ALIZADEHRAD2018303}
D.~Alizadehrad and D.~A. Fedosov,
\newblock Static and dynamic properties of smoothed dissipative particle
  dynamics, Journal of Computational Physics {\bf 356}, 303  (2018).

\bibitem{doi:10.1680/geot.1979.29.1.47}
P.~A. Cundall and O.~D.~L. Strack,
\newblock A discrete numerical model for granular assemblies,
  G^^c3^^a9otechnique {\bf 29}, 47 (1979),
  https://doi.org/10.1680/geot.1979.29.1.47.

\bibitem{koshizuka2018moving}
S.~Koshizuka, K.~Shibata, M.~Kondo, and T.~Matsunaga,
\newblock {\em Moving Particle Semi-implicit Method: A Meshfree Particle Method
  for Fluid Dynamics} (Academic Press, 2018).

\bibitem{doi:10.1080/21664250.2018.1436243}
H.~Gotoh and A.~Khayyer,
\newblock On the state-of-the-art of particle methods for coastal and ocean
  engineering, Coastal Engineering Journal {\bf 60}, 79 (2018),
  https://doi.org/10.1080/21664250.2018.1436243.

\bibitem{SUN2013147}
X.~Sun, M.~Sakai, and Y.~Yamada,
\newblock Three-dimensional simulation of a solid-liquid flow by the dem-sph
  method, Journal of Computational Physics {\bf 248}, 147  (2013).

\bibitem{HE2018548}
Y.~He {\em et~al.},
\newblock A gpu-based coupled sph-dem method for particle-fluid flow with free
  surfaces, Powder Technology {\bf 338}, 548  (2018).

\bibitem{PhysRevE.57.2930}
P.~Espa\~nol,
\newblock Fluid particle model, Phys. Rev. E {\bf 57}, 2930 (1998).

\bibitem{doi:10.1063/1.1711295}
D.~W. Condiff and J.~S. Dahler,
\newblock Fluid mechanical aspects of antisymmetric stress, The Physics of
  Fluids {\bf 7}, 842 (1964),
  https://aip.scitation.org/doi/pdf/10.1063/1.1711295.

\bibitem{lucy1977numerical}
L.~B. Lucy,
\newblock A numerical approach to the testing of the fission hypothesis, The
  astronomical journal {\bf 82}, 1013 (1977).

\bibitem{quarteroni2015modeling}
A.~Quarteroni,
\newblock {\em Modeling the heart and the circulatory system}volume~14
  (Springer, 2015).

\bibitem{Imoto2019}
Y.~Imoto, S.~Tsuzuki, and D.~Nishiura,
\newblock Convergence study and optimal weight functions of an explicit
  particle method for the incompressible navier--stokes equations,
  Computational Particle Mechanics {\bf 6}, 671 (2019).

\bibitem{COLAGROSSI2003448}
A.~Colagrossi and M.~Landrini,
\newblock Numerical simulation of interfacial flows by smoothed particle
  hydrodynamics, Journal of Computational Physics {\bf 191}, 448  (2003).

\bibitem{fischer2000introduction}
A.~C. Fischer-Cripps, E.~F. Gloyna, and W.~H. Hart,
\newblock {\em Introduction to contact mechanics}volume 221 (Springer, 2000).

\bibitem{schwarzkopf2011multiphase}
J.~D. Schwarzkopf, M.~Sommerfeld, C.~T. Crowe, and Y.~Tsuji,
\newblock {\em Multiphase flows with droplets and particles} (CRC press, 2011).

\bibitem{keyfitz2001mathematical}
B.~L. Keyfitz {\em et~al.},
\newblock Mathematical properties of nonhyperbolic models for incompressible
  two-phase flow,
\newblock in {\em Proc. Fourth Int. Conf. Multiphase Flow}, EE Michaelides,
  ICMF, 2001.

\bibitem{doi:10.1021/ie50474a011}
S.~Ergun and A.~A. Orning,
\newblock Fluid flow through randomly packed columns and fluidized beds,
  Industrial \& Engineering Chemistry {\bf 41}, 1179 (1949),
  https://doi.org/10.1021/ie50474a011.

\bibitem{10003815634}
C.~Y. WEN,
\newblock Mechanics of fluidization, Chem. Eng. Prog., Symp. Ser. {\bf 62}, 100
  (1966).

\bibitem{10.5555/76990}
M.~P. Allen and D.~J. Tildesley,
\newblock {\em Computer Simulation of Liquids} (Clarendon Press, USA, 1989).

\bibitem{doi:10.1080/19648189.2008.9693050}
S.~Luding,
\newblock Introduction to discrete element methods, European Journal of
  Environmental and Civil Engineering {\bf 12}, 785 (2008),
  https://doi.org/10.1080/19648189.2008.9693050.

\bibitem{SHAKIBAEINIA201213}
A.~Shakibaeinia and Y.-C. Jin,
\newblock Mps mesh-free particle method for multiphase flows, Computer Methods
  in Applied Mechanics and Engineering {\bf 229-232}, 13  (2012).

\bibitem{GREST1989269}
G.~S. Grest, B.~D^^c3^^bcnweg, and K.~Kremer,
\newblock Vectorized link cell fortran code for molecular dynamics simulations
  for a large number of particles, Computer Physics Communications {\bf 55},
  269  (1989).

\bibitem{10.1145/369534.369540}
T.~Nishimura,
\newblock Tables of 64-bit mersenne twisters, ACM Trans. Model. Comput. Simul.
  {\bf 10}, 348^^e2^^80^^93357 (2000).

\bibitem{10.1145/272991.272995}
M.~Matsumoto and T.~Nishimura,
\newblock Mersenne twister: A 623-dimensionally equidistributed uniform
  pseudo-random number generator, ACM Trans. Model. Comput. Simul. {\bf 8},
  3^^e2^^80^^9330 (1998).

\bibitem{doi:10.1063/1.1883163}
J.~A. Backer, C.~P. Lowe, H.~C.~J. Hoefsloot, and P.~D. Iedema,
\newblock Poiseuille flow to measure the viscosity of particle model fluids,
  The Journal of Chemical Physics {\bf 122}, 154503 (2005),
  https://doi.org/10.1063/1.1883163.

\bibitem{LEI2017571}
H.~Lei, X.~Yang, Z.~Li, and G.~E. Karniadakis,
\newblock Systematic parameter inference in stochastic mesoscopic modeling,
  Journal of Computational Physics {\bf 330}, 571  (2017).

\bibitem{doi:10.1146/annurev.fl.25.010193.000245}
S.~P. Sutera and R.~Skalak,
\newblock The history of poiseuille's law, Annual Review of Fluid Mechanics
  {\bf 25}, 1 (1993), https://doi.org/10.1146/annurev.fl.25.010193.000245.

\bibitem{10.2307/41254479}
P.~H. McGauhey,
\newblock Theory of sedimentation, Journal (American Water Works Association)
  {\bf 48}, 437 (1956).

\bibitem{PhysRevE.97.032611}
N.~Oyama, K.~Teshigawara, J.~J. Molina, R.~Yamamoto, and T.~Taniguchi,
\newblock Reynolds-number-dependent dynamical transitions on hydrodynamic
  synchronization modes of externally driven colloids, Phys. Rev. E {\bf 97},
  032611 (2018).

\end{thebibliography}







\end{document}